\documentclass[10pt]{wlscirep}

\usepackage[numbers,square,comma,sort,compress]{natbib}

\title{Closed-Form Information Capacity of Canonical Signaling Models}

\newcommand{\eg}{{\it e.g.}}

\author[1,*]{Micha\l \ Komorowski}
\affil[1]{Institute of Fundamental Technological Research, Polish Academy of Sciences, Warsaw, Poland}
\affil[*]{m.komorowski@sysbiosig.org}

\begin{abstract}
We employ a unified framework for computing the information capacity of biological signaling systems using Fisher Information. By deriving closed-form or easily computable information capacity formulas, we quantify how well different signaling models, including binomial, multinomial, Poisson, Gaussian, and Gamma distributions, can discriminate among input signals. These expressions clarify how key features such as signal range, noise scaling, pathway length, and receivers' diversity shape the theoretical limits of sensing. In particular, we show how signal-to-noise ratio and fold-change sensitivity arise naturally within the Fisher formalism, and how signal degradation in cascades imposes linear information loss. Our results provide intuitive, analytically grounded tools to benchmark and guide the analysis of real signaling systems, without requiring computationally expensive mutual information estimation. While motivated by cellular communication, the framework generalizes to any system where noisy input–output relationships constrain transmission fidelity, including synthetic biology, sensor networks, and engineered communication channels.
\end{abstract}

\begin{document}

\flushbottom
\maketitle
\thispagestyle{empty}

\section{Introduction}

\noindent 
Cellular signaling systems are fundamental components of living organisms, enabling cells to sense, respond to, and adapt to changing environmental conditions. These pathways support core biological processes including cellular differentiation, immune defense, metabolic regulation, and organismal development \cite{yuan2024guide, klauer2024functional,bray2025modes}. Reliable and accurate signal transmission is vital, as disruptions in signaling can lead to inappropriate cellular responses, compromising fitness and contributing to disease \cite{madsen2025oncogenic, karolak2021concepts,lavoie2020erk}. Consequently, quantifying the precision and robustness of signaling systems has become a tangible effort of modern biosciences \cite{uda2016analysis,kirby2023proofreading,topolewski2021information,tkavcik2025information,
xiong2025quantifying, burt2023distribution,sheu2023quantifying,bedoui2020emerging}.\\

\noindent
Historically, analyses of signaling focused on qualitative descriptions and biochemical mechanisms. More recently, information theory has offered a rigorous, quantitative framework for treating biological signaling systems as noisy communication channels \cite{cover2012elements, araz2023ratio, kuscu2023adaptive, wang2024wavelet, Nurse:2008en}. Within this framework, the concept of information capacity, the maximum amount of information that can be reliably transmitted through a noisy system, has emerged as particularly useful.  \\

\noindent
Formally, in the language of information theory \cite{cover2012elements,polyanskiy2025information,mackay2003information}, a signaling system is described by a conditional distribution \( P(Y \mid X = x) \), representing the likelihood of observing an output \( Y \) given input \( x \). In biological contexts, \( X \) typically denotes a signal such as ligand concentration or stimulus intensity, while \( Y \) represents downstream outputs such as receptor activity, gene expression, or cellular phenotypes. The accuracy of signal transmission depends on sensitivity and noise properties of this mapping as well as the frequency of input presentations, as frequent presentation of inputs that are well discriminated contributes more information than frequent presentation of these poorly discriminated.\\

\noindent
Quantitatively, the amount of information transferred through a signaling system with input distribution \( P(X) \) is measured by the mutual information (MI)
\begin{equation} \label{eq:mi-def}
   MI(X,Y) = \int_{\mathcal X }\int_{\mathcal Y} P(x,y)\log_2\frac{P(y|x)}{P(y)}dx dy,
\end{equation}
where \( P(y) = \int_{\mathcal X} P(y|x)P(x)dx \) is the marginal distribution of outputs, and \( \mathcal X \), \( \mathcal Y \) are the domains of \( X \) and \( Y \), respectively. Mutual information captures the extent to which outputs distinguish inputs, increasing when responses are reproducible and distinct, and decreasing when responses are noisy or overlapping.\\

\noindent
Since mutual information depends on the chosen input distribution \( P(X) \), a more general and interpretable measure is the {information capacity} \( C^* \), defined as the maximum mutual information over all possible input distributions
\begin{equation} \label{eq:cap-def}
    C^* = \max_{P(X)} MI(X,Y).
\end{equation}
The distribution \( P^*(X) \) that achieves this maximum is known as the optimal input distribution and defines the theoretical upper limit of how well a signaling system can resolve different inputs. According to Shannon’s coding theorem \cite{cover2012elements}, this capacity directly corresponds to the maximum number of input states that can be reliably distinguished, e.g., 2 bits of capacity imply discrimination of up to \( 2^2 = 4 \) distinct inputs with negligible error.\\

\noindent
Despite its conceptual power, calculating capacity in practice is challenging. Standard numerical methods such as the Blahut–Arimoto algorithm \cite{blahut1972computation, Arimoto:1972eu,jetka2019information} are computationally intensive and typically require discretization of both input and output spaces, an obstacle for many biological applications. Consequently, there is a need for simplified, analytical tools that make information-theoretic analyses more broadly applicable.\\

\noindent
To address this need, we recently introduced a framework for estimating information capacity based on Fisher Information \cite{jetka2018information}, a core concept in statistical estimation theory \cite{le2012asymptotic,Kass:J8S5SGtR,Clarke:1994gw, Bernardo:1979vc}. Building on this prior work, the present study extends the Fisher Information framework by deriving closed-form expressions for the asymptotic information capacity \( C^*_A \) across several canonical probability distributions frequently used to model biological data, including binomial, multinomial, Poisson, Gaussian, and Gamma distributions.\\

\noindent
These explicit formulas enable rapid estimation of capacity across a range of models, helping to uncover how signaling precision depends on key system parameters. We illustrate these results with canonical biological  examples and provide computational scripts that facilitate application of the theory to other signaling systems. Taken together, our framework aims to bridge the gap between theoretical insight and practical implementation in the analysis of biological information processing.

\section{Materials and Methods}
\noindent
Calculating the information capacity of signaling systems typically involves optimizing 
mutual information, often using numerical methods such as the Blahut–Arimoto algorithm 
\cite{Arimoto:1972eu,blahut1972computation,jetka2019information}. While accurate, these methods 
can be computationally intensive and require discretization of the input and output spaces, 
which becomes challenging, especially for high-dimensional or time-resolved data. 
To address these limitations, we recently proposed an alternative framework based on Fisher Information \cite{jetka2018information}, which we outline below.

\subsection{Fisher Information and Asymptotic Capacity}

\noindent
Consider a signaling system in which the output is denoted by $Y_N = (Y^{(1)}, \dots, Y^{(N)})$,
representing $N$ independent and identically distributed (i.i.d.) observations of a random variable $Y \sim P(\cdot \mid X = x)$. In biological settings, $N$ may correspond to the number of cells, receptor copies, or other parallel signal receivers independently sensing the same input signal $X$.

\noindent
The corresponding information capacity quantifies the maximum mutual information between the input and the full output vector:
\begin{equation}\label{eq:C_N}
C^*(X, Y_N) = \max_{P(X)} I(X; Y_N),
\end{equation}
where $C^*(X, Y_N)$ is the total capacity of the ensemble of $N$ independent systems. The independence assumption holds when the noise sources in each receiver are uncorrelated.

\noindent
To avoid computationally expensive optimization over $P(X)$, we approximate $C^*(X, Y_N)$ using an asymptotic relation grounded in Bayesian statistics and the theory of reference priors \cite{Clarke:1994gw, Bernardo:1979vc, walker1969asymptotic}. This approximation involves the Fisher Information (FI) matrix, denoted $\mathrm{FI}_{ij}(x)$:
\begin{equation} \label{eq:FIM}
\mathrm{FI}_{ij}(x) = 
\mathbb{E}\left[
    \frac{\partial \log P(Y \mid X = x)}{\partial x_i}
    \cdot
    \frac{\partial \log P(Y \mid X = x)}{\partial x_j}
\right],
\end{equation}
where $x = (x_1, \dots, x_l)$ and $l$ is the dimension of the input space.

\noindent
Under regularity conditions, one obtains the following asymptotic relation:
\begin{equation} \label{eq:C_N_asymptote}
C^*(X, Y_N) - \frac{l}{2} \log_2 N \;\xrightarrow[N \to \infty]{}\; C^*_A,
\end{equation}
where \( C^*_A \) is known as the {asymptotic information capacity} and has the closed-form:
\begin{equation} \label{eq:CapacityApp}
C^*_A = \log_2\left( (2\pi e)^{-l/2} V \right),
\end{equation}
with
\begin{equation} \label{eq:V}
V = \int_{\mathcal{X}} \sqrt{ |\mathrm{FI}(x)| }\, \mathrm{d}x,
\end{equation}
where $|\cdot|$ denotes the determinant.

\noindent
From this, we define the approximation:
\begin{equation} \label{eq:CapacityN}
C^*_N = C^*_A + \frac{l}{2} \log_2 N,
\end{equation}
so that
\begin{equation}
C^*(X, Y_N) - C^*_N \;\xrightarrow[N \to \infty]{}\; 0.
\end{equation}

\subsection{Interpreting \( C^*(X, Y_N) \), \( V \), \( C^*_A \), and \( C^*_N \)}

\noindent
To clarify the interpretation of key quantities used in the derivation, we provide a concise summary of their meaning and relevance.

\medskip
\noindent
The quantity \( C^*(X, Y_N) \) denotes the {true information capacity} of a system composed of \( N \) independent receivers, each observing the same input signal \( X \). It is defined as the maximum mutual information between \( X \) and the joint response vector \( Y_N = (Y^{(1)}, \dots, Y^{(N)}) \), optimized over all input distributions. This is the theoretical gold standard for quantifying information transmission in a population of sensors.

\medskip
\noindent
The term \( V = \int_{\mathcal{X}} \sqrt{|\mathrm{FI}(x)|} \, dx \), Fisher information based sensitivity metric, captures the total {resolvability} of the input space \( \mathcal{X} \). Here, \( \sqrt{|\mathrm{FI}(x)|} \) reflects the local density of distinguishable input values near point \( x \), as defined by the Fisher Information metric. In one dimension, \( V \) corresponds to the length of the input space measured in statistical units,  analogous to a Riemannian arc length in information geometry. As such, \( V \) quantifies how finely the input range can be partitioned into reliably distinguishable regions, given the system's noise and sensitivity.

\medskip
\noindent
The quantity \( C^*_A \) transforms this geometric quantity \( V \) into Shannon bits and represents the {asymptotic information capacity} of the system. It is the leading-order term in the large-\( N \) representation of \( C^*(X, Y_N) \) and depends only on the structure of the conditional distribution \( P(Y \mid X) \), not on any particular input distribution or the number of sensors. Intuitively, it sets the baseline capacity contributed by a single receiver to the ensemble total, establishing the intercept in the scaling law for \( N \) independent observations.

\medskip
\noindent
The capacity \(C^*_N \) captures the asymptotic scaling behavior of the capacity for a population of \( N \) independent sensors. It provides a computationally efficient alternative to direct numerical optimization of mutual information, and converges to \( C^*(X, Y_N) \) as \( N \) increases. This formula is especially useful because it replaces costly mutual information optimization with a simple integral of Fisher Information. 

\medskip
\noindent
While \( C^*_A \) is defined independently of \( N \), we note that for \( N = 1 \) this formula yields \( C^*_1 = C^*_A \). As such, it can also serve as a practical first-order surrogate for the single-receiver capacity 
\( C^*_1 \approx C^*(X, Y) \). However, this identification is not generally exact, and care must be taken in interpreting \( C^*_1 \) as \( C^*(X, Y) \) without additional validation. Still, the approximation is well justified when the output variable \( Y \) can be represented as a sufficient statistic for a larger ensemble of independent observations (due to equivalency of likelihoods),  for example, when \( Y \) summarizes many identically distributed auxilliary variables, e.g., Gaussian variable as a mean of other Gaussian variables. Moreover, the approximation \( C^*_1 \approx C^*(X, Y) \) is intuitively expected to hold in regimes where the Fisher Information \( \mathrm{FI}(x) \) is large and smoothly varying across the domain. In such cases, the llikelihood behaves locally like a well-resolved, low-noise Gaussian model, and the asymptotic expression should captures the essential geometry of signal discrimination even at small \( N \).
\subsection{When does the approximation hold?}
\noindent 
The Fisher-based asymptotic approach, stemming from  on convergence in Eq.  \ref{eq:C_N_asymptote}, relies on statistical regularity and non-singularity assumptions to ensure accurate capacity estimates under large-$N$ conditions. The requirements are as follows:

\begin{enumerate}
  \item {Regularity Conditions.}
    \begin{itemize}
      \item {Smoothness:} The likelihood function $P(Y \mid X = x)$ and its log should be
            sufficiently differentiable with respect to $x$. This
            allows the Fisher Information $\mathrm{FI}(x)$ to be well-defined and continuous.
      \item {Identifiability:} If distinct parameter values $x \neq x'$ yield the
            same output distribution, the inference of $x$ is an ill-posed problem as the model cannot distinguish between these values.
      \item {Finite Integrals:} Integrals involving $\sqrt{\lvert \mathrm{FI}(x)\rvert}$
            should converge over the input domain $\mathcal{X}$. 
    \end{itemize}

  \item {Non-singularity of the Fisher Information Matrix.}
    If $|\mathrm{FI}(x)|=0$ for some $x \in \mathcal{X}$, at least one direction in the input
    space fails to change the output distribution. 

  \item {Independence of Receivers.}
    The outputs $Y^{(1)}, \dots, Y^{(N)}$ must be independent for the joint
    distribution to factorize as $\prod_{i=1}^{N} P(Y^{(i)} \mid X)$. 

\end{enumerate}
\noindent 
In practice, most commonly studied biochemical system models satisfy these regularity criteria, especially
when the input-output mapping $P(Y \mid X = x)$ is smoothly parameterized and each dimension of
$x$ exerts a distinct effect on $Y$. For in-depth mathematical treatments, see
\cite{Clarke:1994gw,Bernardo:1979vc} and references therein.

\subsection{Negative Values of $C_{\mathrm{A}}^*$}

\noindent
Although the single-receiver capacity \( C(X,Y)^* \) is guaranteed to be non-negative, the {asymptotic capacity} \( C_{\mathrm{A}}^* \) can, in some cases, take negative values. This situation arises when the system's actual ability to distinguish between inputs grows only very slowly with the number of receivers \( N \). In such cases, even large populations of receivers accumulate little additional information.
\noindent
This outcome is not a flaw of the approach but a necessary consequence of the scaling behavior implied by Eqs.~\ref{eq:C_N_asymptote} and \ref{eq:CapacityN}. If \( C_{\mathrm{A}}^* \) were  constrained to be non-negative, it would incorrectly imply that every system achieves at least \( \frac{l}{2} \log_2(N) \) bits of capacity for large \( N \), a bound that could be clearly violated. For further interpretation, consider two signaling systems: one with \( C_{\mathrm{A}}^* = -1 \) bit and another with \( C_{\mathrm{A}}^* = 1 \) bit. As \( N \) increases, the first system's capacity \( C_N^* \) remains exactly 2 bits below that of the second and its number of resolvable inputs $2^{C^*_N}$ grows at only one-fourth the rate.
\noindent
In practical settings, when the aim is to estimate the single-receiver capacity \( C^*(X,Y) \) and the asymptotic capacity \( C_{\mathrm{A}}^* \) turns out to be negative, we recommend switching to a direct method, such as numerical integration or the Blahut--Arimoto algorithm. These approaches avoid the asymptotic approximation and always yield a strictly nonnegative estimate, albeit at the cost of higher computational effort.

\subsection{Handling Parameter Transformations}
\noindent
Frequently, the distribution of interest \eg, Gaussian, is expressed in terms of a canonical
parameter vector ${\theta} \in \mathbb{R}^{l'}$, rather than the signal
$x \in \mathbb{R}^l$ directly. Suppose ${\theta} = h(x)$ is a smooth map
from $\mathbb{R}^l \to \mathbb{R}^{l'}$. Then, if $\mathrm{FI}({\theta})$
denotes the Fisher Information with respect to ${\theta}$, the
FI with respect to $x$ is given by
\begin{equation}\label{eq:FiTrans}
  \mathrm{FI}(x) 
  \;=\; 
  J(x)^\mathsf{T}\,\mathrm{FI}({\theta})\,J(x),
  \quad\text{where}\quad
  J(x) \;=\;\frac{\partial {\theta}}{\partial x}
  \;\in\;\mathbb{R}^{{l'} \times l}.
\end{equation}
Here, $J(x)$ is the Jacobian matrix of partial derivatives
\[
J(x)\;=\;
\begin{bmatrix}
  \tfrac{\partial \theta_1}{\partial x_1} & \cdots & \tfrac{\partial \theta_1}{\partial x_l}\\
  \vdots & \ddots & \vdots \\
  \tfrac{\partial \theta_{l'}}{\partial x_1} & \cdots & \tfrac{\partial \theta_{l'}}{\partial x_l}
\end{bmatrix}.
\]
Thus, one can compute $\mathrm{FI}({\theta})$ in whichever parameterization
is most straightforward (e.g.\ if ${\theta}$ has a known closed-form expression),
and then transform it back into $\mathrm{FI}(x)$ by applying Eq. \ref{eq:FiTrans}.

\noindent 
As an example, it is worth considering log-parametrization. In many signaling contexts, inputs $x$ range over several orders of magnitude (e.g.\ ligand concentrations). It can be numerically more stable to define $z = \log_{10}(x)$ and work
with $\mathrm{FI}(z)$. The Jacobian here is
\[
  \frac{\mathrm{d} x}{\mathrm{d} z} 
  \;=\; x\,\ln(10),
\]
so $\mathrm{FI}(z) = \mathrm{FI}(x)\,\bigl(x\,\ln(10)\bigr)^2$.

%%%%%%%%%%%%%%%%%%%%%%%%%%%%%%%%%%%%%%%%%%
\section{Results}

\noindent
In this section, we apply the Fisher Information-based framework from the Methods to derive closed-form expressions for the information capacity of systems modeled by several canonical probability distributions and associated signaling models. These distributions naturally emerge in well-defined biochemical contexts and cover both discrete and continuous signal-response relationships. Their analytical tractability allows us to pinpoint how key biological parameters, such as ligand binding affinities, signal intensities, transition rates, and noise scaling, shape information transmission. In contrast to simulation-based approaches, our method provides explicit, interpretable formulas that enable rapid capacity estimation and support both theoretical understanding and model-guided circuit design.

\subsection{Binomial and Bernoulli Distribution}

\noindent
Many signaling systems produce binary responses: a biochemical sensor (receptor) or a transcription factor is either bound or unbound, ON/OFF, depending on input. These systems can be modeled using  binomial or Bernoulli processes. One of the classic examples of such a sensor is the \textit{E.~coli} AraC protein, which senses the presence or absence of 
L-arabinose and adopts either a repressing or activating conformation. In the absence of the ligand 
(OFF), downstream genes remain silent, whereas in the presence of L-arabinose (ON), AraC promotes 
transcription \cite{schleif2000regulation}. This kind of binary response is well captured by 
binomial/Bernoulli modeling. 

\noindent
Formally, for a single Bernoulli trial where the random variable \( Y \in \{0, 1\} \) equals 1 with probability \( p \), the Fisher Information with respect to \( p \) is
\begin{equation}
FI(p) = \frac{1}{p(1 - p)}.
\end{equation}

\noindent
This expression quantifies how sensitively the likelihood responds to changes in \( p \).  Using Eq.~\ref{eq:V}, we obtain
\begin{equation}\label{eq:V_Binomial}
V = \int_{p_{\min}}^{p_{\max}} \frac{1}{\sqrt{p(1 - p)}} \, dp = \arcsin(2p_{max}-1)-\arcsin(2p_{min}-1),
\end{equation}

\noindent leading to the asymptotic capacity as defined by Eq.~\ref{eq:CapacityApp}

\begin{equation}
C_{\text{A}}^*= \log_2\left({(2\pi e)^{{-}\frac{1}{2}}} \left(\arcsin(2p_{max}-1)-\arcsin(2p_{min}-1) \right) \right).
\end{equation}

\noindent For $N$ independent Bernoulli copies $Y^{(1)},\dots,Y^{(N)}$, according to Eq. \ref{eq:CapacityN} the capacity can be approximated as  

\begin{equation}
C^*_N = \frac{1}{2}\log_2\left(\frac{N \cdot \left[\arcsin(2p_{max}-1)-\arcsin(2p_{min}-1)\right]^2}{{2e\pi}} \right),
\end{equation}

\noindent which simplifies further if $p_{\min}=0$ and $p_{\max}=1$

\begin{equation}\label{eq:CapacityN_simplified}
C^*_N = \frac{1}{2} \log_2\left(\frac{N\pi}{2e}\right).
\end{equation}

\noindent 
The above expression is the information capacity of the Bernoulli distribution. 
To ensure statistical rigor in transforming the capacity of \( N \) independent binomial 
variables, \(Y^{(1)}, \dots, Y^{(N)}\), into the capacity of a Bernoulli variable, $\sum_{i=1}^N Y^{(i)} $,  we note that the two formulations 
are equivalent because the joint vector \( Y_N = (Y^{(1)}, \dots, Y^{(N)}) \) and the 
aggregate count \( \sum_{i=1}^N Y^{(i)} \) contain the same amount of information about 
the underlying parameter \( p \). This is due to the fact that \( \sum_{i=1}^N Y^{(i)} \) 
is a sufficient statistic for \( p \) under the binomial model.

\subsubsection*{Example: Simple Biochemical Sensor}

\noindent Consider a system of \( N \) identical biochemical receptors, each of which is active with probability \( p = h(x) \), where \( x \) denotes the concentration of an activating ligand, e.g., arabinose in the case of the AraC protein. A typical functional form for \( h(x) \) is the Hill function \cite{cornish2013fundamentals}
\begin{equation}
 h(x) = \frac{(x/H)^n}{1 + (x/H)^n},
\end{equation}
where \( H \) is the half-maximal effective concentration, and \( n \) is the Hill coefficient, which quantifies the cooperativity of ligand binding, i.e., how binding at one site influences affinity at others. For non-cooperative systems, \( n = 1 \). Using the variable transformation formula, Eq. \ref{eq:FiTrans}, we get Fisher information

\begin{equation}
FI(x) = \frac{1}{{h(x)(1 - h(x))}} \left( \frac{\partial h(x)}{\partial x} \right)^2.
\end{equation}

\noindent The corresponding Fisher-based sensitivity metric, Eq. \ref{eq:V}, required for capacity calculation is
\begin{equation}
V = \int_{x_{\min}}^{x_{\max}} \frac{1}{\sqrt{h(x)(1 - h(x))}} \left| \frac{\partial h(x)}{\partial x} \right|\, dx = \int_{h\left(x_{\min}\right)}^{h\left(x_{\max}\right)} \frac{1}{\sqrt{p(1 - p)}} \, dp,
\end{equation}
where \( x_{\min} \) and \( x_{\max} \) specify the operational signal range. This integral quantifies the system’s ability to distinguish between different input concentrations. Interestingly, the value of \( V \) depends only on the endpoints \( h(x_{\min}) \) and \( h(x_{\max}) \), and not on the detailed shape or steepness of \( h(x) \) in between \cite{komorowski2019limited, martins2011trade}. 

\noindent For example, if \( h(x) \) transitions fully from 0 to 1 over the signal range, i.e., \( p_{\min} = 0 \), \( p_{\max} = 1 \), according to Eq. \ref{eq:V_Binomial}, we have $V=\pi$ resulting in the capacity
\begin{equation}\label{eq:CapacityN_simplified}
C^*_N = \frac{1}{2} \log_2\left(\frac{N \pi}{2e}\right).
\end{equation}

\noindent This result illustrates that systems that span the full output range \([0,1]\) have the same  information capacity regardless of the slope of the transition. Thus, a steep response curve does not  increase capacity, perhaps contrary to intuition. The framework here exposes such not necessarily obvious non-obvious dependencies of capacity on model parameters, highlighting the importance of range, and not  sensitivity, in optimizing signal transmission in binomial systems \cite{komorowski2019limited, martins2011trade}.

\subsection{Multinomial Distribution}
\noindent In many biological signaling systems, a single sensor does not merely toggle between two states but instead produces multiple discrete outcomes, each associated with a distinct functional consequence. Such behavior can  arise from multi-state receptors and corresponding downstream processing modules resulting in more than two  outputs. 

\noindent
A canonical example is the $\beta_2$-adrenergic receptor, a G-protein-coupled receptor (GPCR) that can adopt several active conformations. These states selectively engage different intracellular partners, such as G-proteins or arrestins, thereby initiating distinct signaling cascades \cite{wisler2014recent, Kenakin:2016vn}. These different signaling outcomes are naturally modeled using the multinomial distribution, which generalizes the binomial framework to account for systems with more than two discrete response states.

\noindent In what follows, we show how to compute the information capacity for such multinomial signaling systems and how the structure of their response profiles, e.g., overlap or separation between response modes, modulates information transmission.

\noindent Formally, in the multinomial setting, a system performs $N$ independent `trials' each yielding
one of $m$ distinct outcomes with probabilities $p_1, p_2, \dots, p_m$, where
$\sum_{j=1}^m p_j = 1$. Let $Y = (Y^1, Y^2, \dots, Y^m)$ denote the vector of counts, so that $Y^j$
is the number of trials that end up in outcome $j$. Formally,
\[
  Y \;=\; (Y^1, Y^2, \dots, Y^m)
  \;\sim\;
  \text{Multinomial}\bigl(N,\;p_1, p_2,\dots,p_m\bigr),
\]
and the probability mass function is
\[
  p\bigl(y_1,y_2,\dots,y_m; N, p_1,\dots,p_m\bigr)
  \;=\;
  \frac{N!}{y_1!\,y_2!\cdots y_m!}\;
  p_1^{\,y_1}\,
  p_2^{\,y_2}\cdots
  p_m^{\,y_m},
\]
subject to $\sum_{j=1}^m y_j = N$ and $p_j\ge 0$. 

\noindent
 The Fisher Information for the Multinomial$(N,p_1,\dots,p_m)$
distribution with a parameter vector $p$ takes a particularly simple diagonal form

\begin{equation}
FI({p}) = N
\begin{bmatrix}
\frac{1}{p_1} & 0 & \cdots & 0 \\
0 & \frac{1}{p_2} & \cdots & 0 \\
\vdots & \vdots & \ddots & \vdots \\
0 & 0 & \cdots & \frac{1}{p_m}
\end{bmatrix}.
\end{equation}

\noindent
Because the probabilities $p_1, p_2, \dots, p_m$ must sum to 1, only $m-1$ of them are truly 
independent.  The resulting singularity of Fisher information implies that additional assumptions or context 
regarding how input signal drives each $p_j$ are needed to quantify information transfer. As a concrete 
illustration, we next examine an example of an {allosterically regulated sensor}, 
where different ligand-binding states map onto multiple output states.

\subsubsection*{Allosterically Regulated Sensor}
\noindent Beyond the GPCR example \cite{wisler2014recent,Kenakin:2016vn}, many sensory systems exhibit multi-state allosteric regulation.
Certain enzymes, such as aspartate transcarbamoylase (ATCase) or phosphofructokinase‐1 (PFK‐1), respond 
to multiple effector ligands (e.g., nucleotides or substrates) and can occupy distinct ligand-bound states 
with varying catalytic activities \cite{cornish2013fundamentals,olsman2016allosteric,martins2011trade}. Another prominent example is the epidermal growth factor receptor (EGFR), a receptor tyrosine kinase that binds a variety of ligands, including EGF. These ligands differentially stabilize receptor conformations and lead to distinct phosphorylation patterns, thereby recruiting different adaptor proteins and selectively activating downstream signaling pathways such as MAPK, PI3K/Akt, or STAT cascades \cite{lemmon2010cell}.
In such systems, varying ligand concentrations shape the probability distribution over multiple conformational states, rather than inducing a simple binary switch.
This naturally motivates the use of multinomial models to capture the multi-outcome signaling behavior

\noindent In mathematical modeling, these  conformations map onto multinomial outcomes  \cite{komorowski2019limited,martins2011trade, olsman2016allosteric,cornish2013fundamentals}.  Here, we assume the sensor can take \( k \) distinct functional states, each associated with a potential of activating a distinct signaling outcome, \eg, distinct signaling pathway. Each sensor molecule  takes the functional state $j$ with probability \( \omega_i \) for \( i = 1,\dots,k \). Probabilities of taking different functional states  are often dependent on concentrations of modulatory ligands, which for tractability is not modeled here.  Each functional state can then be active with probability \( h_i(x) \), where \( x \) is the input signal (such as input ligand concentration), or inactive with probability \( 1 - h_i(x) \). Therefore, overall, the system occupies \( m = 2k \) distinct states, with the probabilities of each state given by

\begin{equation}
p=\left(\omega_1 h_1(x), \omega_1 (1 - h_1(x)), \omega_2 h_2(x), \omega_2 (1 - h_2(x)), \dots, \omega_k h_k(x), \omega_k (1 - h_k(x))\right).
\end{equation}

\begin{figure}[ht]
\centering
\begin{minipage}{\linewidth}
 \flushleft
% {\bf (A)}\\ % Label for the first figure
 \includegraphics[width=\linewidth]{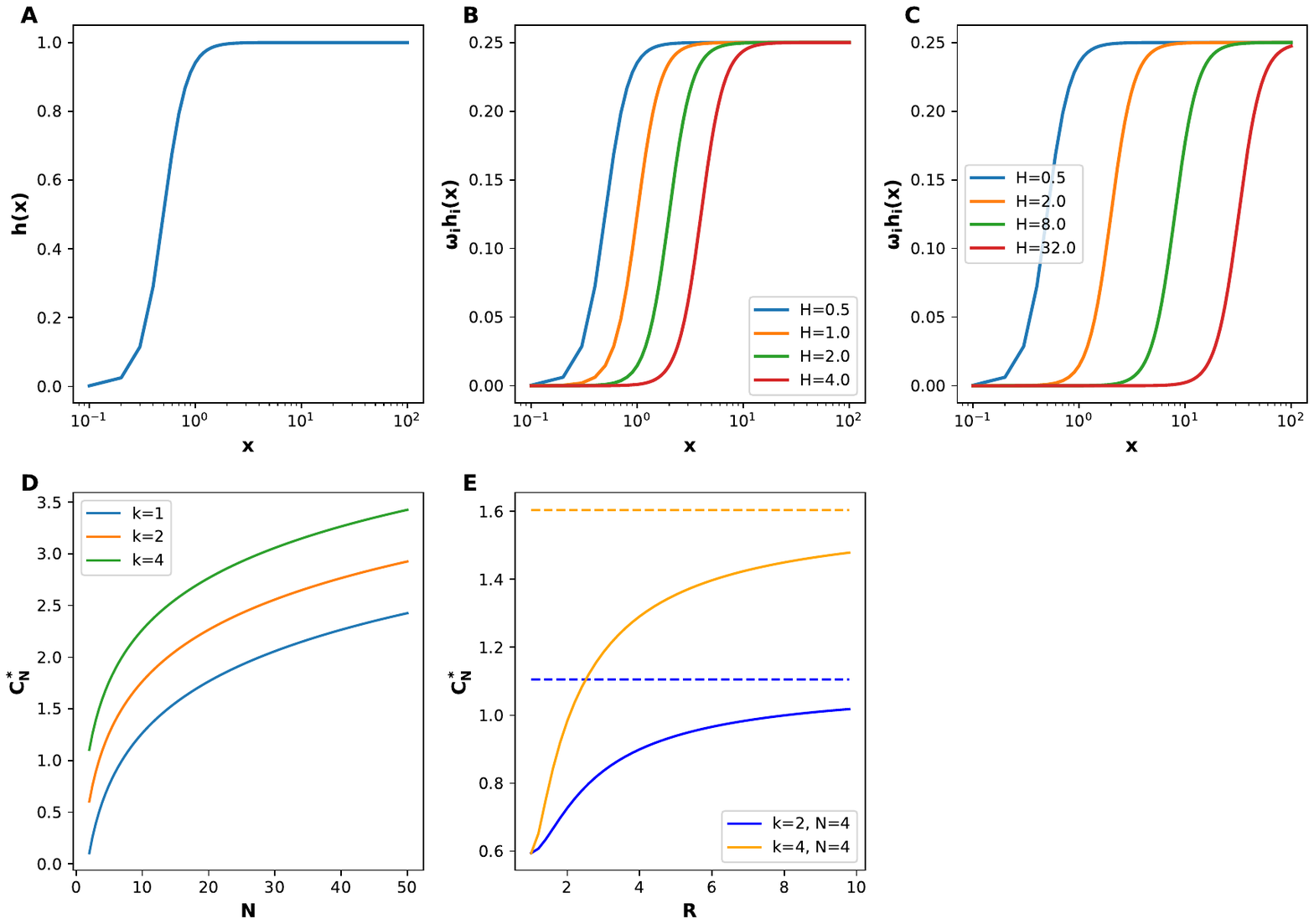}
\end{minipage}
\caption{\ \\
{\bf (A)} The probability of a sensor being active as a function of the signal $x$, modeled as Hill function $h(x)=\frac{(x/H)^n}{1+(x/H)^n}$ with $H=1$, $n=2$.\\
{\bf (B)} The probability of a sensor in each of the four equiprobable ($\omega_i=1/k$) functional states (k=4) being active, $\omega_i h_i(x)$, as a function of the signal $x$ for a strongly overlapping sensitivity ranges of each functional state modeled as a Hill function with $H_1=0.5$, $H_2=1$, $H_3=2$, $H_4=4$.\\
{\bf (C)} Same as in (C) but for small overlaps between sensitivity ranges of each functional state: $H_1=0.5$, $H_2=2$, $H_3=8$, $H_4=32$.\\
{\bf (D)} Capacity $C_N$ as a function of $N$ in a scenario with one conformational state ($k=1$, Binomial output), as well as two and four conformational states (k=2 and k=4, Multinomial output).\\
{\bf (E)} Capacity $C_N$ in a scenario with two $(k=4)$ and four $(k=4)$ conformational states as a function of the ratio $R=H_{i+1} / H_{i}$, together with the limit of fully distinct responses for each functional state established by Eq. \ref{eq:multinomial:capacity:limited}.
}
\label{fig:multinomial}
\end{figure}

\noindent This is visually represented in Figure \ref{fig:multinomial}A-C. Figure \ref{fig:multinomial}A shows the probability of a receptor being active for the case where \( k = 1 \) (a single functional state) as in the Binomial scenario. Panels {B}\&C correspond to a system with \( k = 4 \) functional states, showing how different functional states respond by turning active with increasing ligand concentrations. Panel B illustrates overlapping sensitivity ranges for each functional state, modeled using different values of the Hill parameter \( H \). As the value of \( H \) increases, the sensitivity of each state shifts, allowing the system to respond in a higher range of signal values. The ratio of subsequent Hill parameters $R=H_{i+1}/H_i$ tunes therefore separation of sensitivity ranges of subsequent functional states,  with higher $R$ implying stronger separation. 
In panel B the ratio between subsequent Hill parameters $R=2$ results in moderately overlapping sensitivity ranges. Panel C illustrates the case where the sensitivity ranges are more distinct, which is achieved by increasing the ratio R up to 4. 

\noindent Having defined how  the input signal $x$ maps onto parameter vector $p$, the Jacobian matrix \( J(x) \), which accounts for the transformation of the FI, is given by 

\begin{equation}
 J(x) = \left(\omega_1 \frac{\partial h_1(x)}{\partial x}, -\omega_1 \frac{\partial h_1(x)}{\partial x}, 
 \omega_2 \frac{\partial h_2(x)}{\partial x}, -\omega_2 \frac{\partial h_2(x)}{\partial x}, \dots, 
 \omega_k \frac{\partial h_k(x)}{\partial x}, -\omega_k \frac{\partial h_k(x)}{\partial x}\right),
\end{equation}

\noindent which leads to the Fisher Information for the system

\begin{equation}
FI(x) = \sum_{i=1}^k \omega_i \left( \frac{\partial h_i(x)}{\partial x} \right)^2 \left(\frac{1}{h_i(x)} + \frac{1}{1 - h_i(x)}\right).
\end{equation}

\noindent Numerically, this expression can be evaluated to compute the system's information capacity. However, we can derive a closed-form solution under a three simplifying assumptions. Consider a scenario where the sensitivity ranges of all functional states \( h_i(x) \) are non-overlapping, meaning that for any given signal \( x \), only one of the response functions \( h_i(x) \) has non-zero first derivative \( \frac{\partial h_i(x)}{\partial x} \). Besides, assume even distribution of receptors among all functional states (\( \omega_i = 1/k \)), for $i=1,...,k$.
This assumptions, along with $h_i(0)=0$ and $h_i(\infty)=1$, for $i=1,...,k$, allows us to simplify the integration of the Fisher Information. Integrating the Fisher Information under these assumptions gives
\[
V = \int_0^{\infty} \sqrt{\sum_{i=1}^k \frac{1}{k} \left( \frac{\partial h_i(x)}{\partial x} \right)^2 \left(\frac{1}{h_i(x)} + \frac{1}{1 - h_i(x)}\right)}dx = 
\sqrt{\frac{1}{k}} \sum_{i=1}^k \int_0^{1} \ \left(\frac{1}{\sqrt{q_i (1 - q_i)}}\right)dq_i = \sqrt{k} \pi.
\]

\noindent 
Substituting this result into the capacity equation, we obtain

\begin{equation} \label{eq:multinomial:capacity:limited}
C_N^* = \frac{1}{2} \log_2\left(\frac{k N \pi}{2e}\right).
\end{equation}

\noindent
This result establishes an upper bound on the information capacity achievable when a sensor operates through \(k\) distinct functional states, each tuned to a different segment of the input range. As shown in Figure~\ref{fig:multinomial}D, the capacity \(C_N^*\) grows with both the number of receptors \(N\) and the number of functional states \(k\). The benefit from higher $k$ however depends on how distinct sensitivity ranges of the functional states are. 

\noindent
Further, Figure~\ref{fig:multinomial}E  demonstrates that how closely to the upper bound a system operates depends critically on how well-separated the response profiles \(h_i(x)\) are. When these profiles overlap substantially, i.e., when the Hill parameter ratio \(R = H_{i+1}/H_i\) is small, states become redundant and add little new information. As \(R\) increases and the profiles become more distinct, the system transitions from redundancy-limited to resolution-limited behavior, with capacity approaching the theoretical maximum derived in Eq.~\ref{eq:multinomial:capacity:limited}. This illustrates how quantitative tuning of biochemical parameters governs the efficiency of information transfer.

\noindent
In summary, this example illustrates how the proposed framework can be used to quantify the 
benefits of conformational diversity in allosteric sensors. It provides a tractable way to 
assess whether additional receptor states meaningfully improve signal discrimination, and how this 
depends on the tuning of their sensitivity profiles. Besides,  because the multinomial distribution and its 
Fisher Information generalize naturally to other systems with discrete multi-state outputs, such as 
stochastic biochemical networks modeled via the Chemical Master Equation 
\cite{jahnke2007solving, gadgil2005stochastic}, the same analytical tools can be broadly applied 
across diverse molecular signaling contexts.

\clearpage

\subsection{Poisson Distribution}

\noindent
The Poisson distribution is an essential model for biochemical signaling systems in which outputs are counts of discrete, stochastic events, such as transcription initiation, ion channel openings, receptor–ligand interactions, or molecular arrivals. These processes are typically memoryless and governed by exponential waiting times, making the Poisson model a natural fit. The distribution is parameterized by \( \lambda > 0 \), the expected number of events, which directly captures both the signal strength and its inherent noise level. Furthermore,  the Poisson model connects directly to the structure of the Chemical Master Equation \cite{jahnke2007solving, gadgil2005stochastic}, enabling rigorous, first-principles modeling of information flow through noisy molecular pathways. In the following analysis, we derive exact expressions for the information capacity of Poissonian systems and show how signal range, pathway length, and degradation rates constrain the precision of information transfer.

\noindent
We denote \( Y \sim \text{Poisson}(\lambda) \), meaning that the random variable \( Y \) takes integer 
values \( y = 0, 1, 2, \ldots \) with probability
\[
\text{Poisson}(Y = y \mid \lambda) = f(y; \lambda) = \frac{\lambda^y e^{-\lambda}}{y!}.
\]

\noindent The Fisher Information for the rate parameter \( \lambda \) is

\[
FI(\lambda) = \frac{1}{\lambda}.
\]

\noindent Using the formalism introduced, we integrate the square root of the Fisher Information, Eq. (\ref{eq:V}). Integrating over the range from $\lambda_{\min}$ to $\lambda_{\max}$, we have

\[
V = 2\sqrt{\lambda_{\text{max}}} - 2\sqrt{\lambda_{\text{min}}}.
\]

\noindent Substituting this into asymptotic information capacity formula, Eq.~\ref{eq:CapacityApp}, results in

\begin{equation}
C_{\text{A}}^* = \log_2\left({(2\pi e)^{-\frac{1}{2}}} \left(2\sqrt{\lambda_{\text{max}}} - 2\sqrt{\lambda_{\text{min}}}\right)\right).
\end{equation}

\noindent
Unlike binomial or multinomial models, where the asymptotic regime is naturally defined by the number of trials \( N \), the Poisson distribution lacks an explicit structure representing different receivers. Nevertheless, an analogous asymptotic interpretation remains valid and useful. Consider partitioning a Poisson-distributed variable \( Y \sim \text{Poisson}(\lambda) \) into a sum of \( N \) independent Poisson components 
\[
Y^{(i)} \sim \text{Poisson}\left(\tfrac{\lambda}{N}\right) \quad \text{for } i = 1, \dots, N,
\]
such that \( Y = \sum_{i=1}^N Y^{(i)} \). Since the sum of independent Poisson variables equals a Poisson with summed rates, this decomposition is exact. Moreover, the total count \( \sum_{i=1}^N Y^{(i)} \) is a sufficient statistic for \( \lambda \), meaning that the full collection \( Y^N = (Y^{(1)}, \dots, Y^{(N)}) \) contains no more information about \( \lambda \) than the sum alone.

\noindent
This identity justifies treating \( N \) as an effective number of independent signal receivers, akin to other distributions. Therefore, increasing \( \lambda \) mirrors the the asymptotic behavior of the capacity in terms of increasing \( N \). When the rate parameters of interest are larger, specifically, when \( \lambda_{\min} \) becomes sufficiently large, the asymptotic capacity \( C^*_{\mathrm{A}} \) becomes a tight approximation of the exact capacity \( C^* \):
\begin{equation}\label{eq:poisson_assym}
C^*_{\mathrm{A}} - C^* \xrightarrow[ \lambda_{\min} \rightarrow \infty ]{} 0,
\end{equation}
and hence \( C^*_{\mathrm{A}} \approx C^* \).

\noindent
This result offers practical value: for high-rate Poissonian processes (e.g., frequent transcription, fast molecular turnover), the asymptotic expression can replace costly numerical estimation without significant loss of accuracy, streamlining the analysis of  signal fidelity.

\subsubsection*{Biochemical cascade}

\begin{figure}[ht]
\centering
\begin{minipage}{\linewidth}
 \flushleft
 {\bf A}\\ % Label for the first figure
 \includegraphics[width=\linewidth]{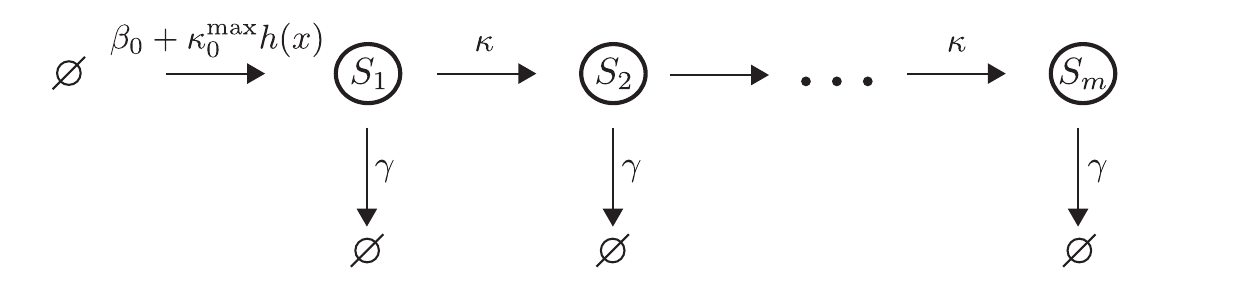}
\end{minipage}

\begin{minipage}{\linewidth}
 \flushleft
 {\bf B}\\ % Label for the second figure
 \includegraphics[width=\linewidth]{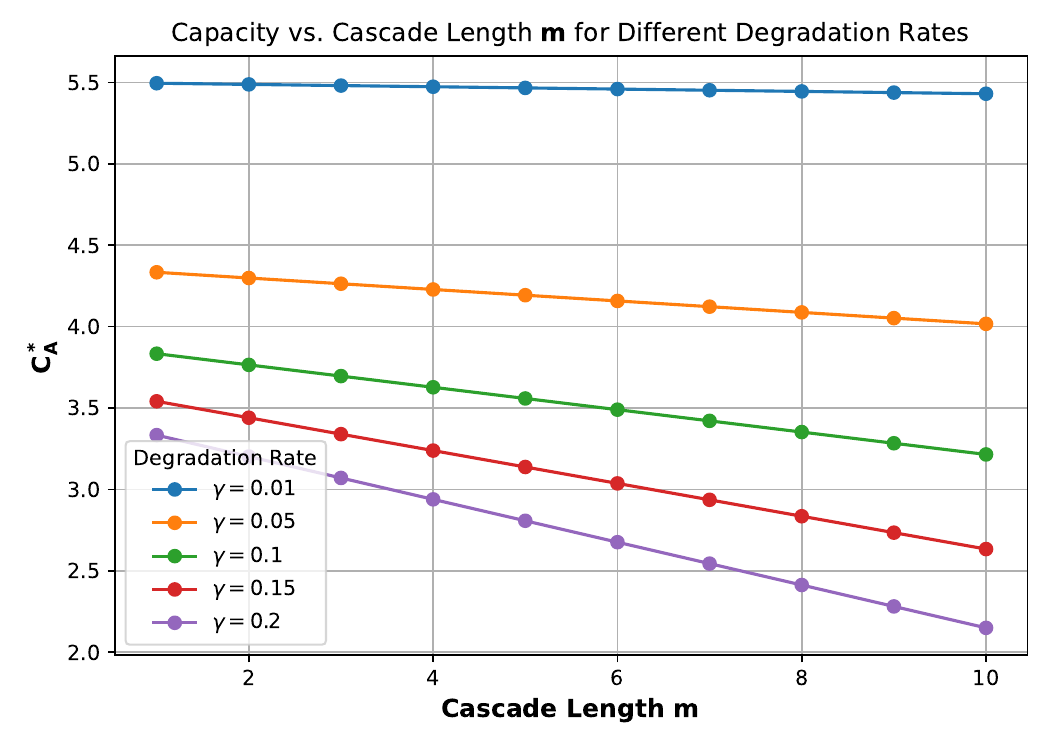}
\end{minipage}

\caption{{\bf (A)} Graphical representation of a chemical linear conversion pathway of length $m$, in which molecules appear in state $S1$ at rate $\kappa_0 = \beta_0+\kappa_0^{\max} \cdot h(x)$ that depends on the external signal $x$ through a monotonic function of choice, here Hill function $h(x)= (x/H)^n/(1+(x/H)^n)$, and then are converted to the next subsequent state at rate $\kappa$ or degrade at rate $\gamma$. All events are assumed to have exponential waiting time implying Poissonian distribution of molecule copy number in each state.\\
{\bf (B)} Information capacity as a function of cascade length for different values of the degradation rate $\gamma$. For computations, we assumed ${\color{black}\beta_0=0.5}$, $\kappa=1$, $\kappa_0^{\max}={\color{black}20}$, $H=1$, $n=2$.} 
\label{fig:cascade}
\end{figure}

\noindent
To illustrate the practical utility of the Poisson capacity formulas, we consider a canonical biochemical cascade composed of \( m \) molecular states \( S_1, S_2, \ldots, S_m \), depicted in Figure~\ref{fig:cascade}A. Molecules enter the initial state \( S_1 \) at a signal-dependent rate
\[
\kappa_0(x) = \beta_0 + \kappa_0^{\max} \cdot h(x),
\]
where \( \beta_0 \) is a constant basal input rate, and \( h(x) = \frac{(x/H)^n}{1 + (x/H)^n} \) is a Hill function modeling activation by an external signal \( x \). Here, \( H \) is the input level for half-maximal activation, and \( n \) determines the steepness (cooperativity) of the input–output response.

\noindent
Molecules transition sequentially from state \( S_i \) to \( S_{i+1} \) at a uniform conversion rate \( \kappa \), while each state also undergoes first-order degradation at rate \( \gamma \). Because transitions and degradation are governed by exponential waiting times, the system dynamics can be modeled as a continuous-time Markov process, and molecule counts at each state follow Poisson distributions \cite{jahnke2007solving, gadgil2005stochastic}.

\noindent
At steady state, balance equations can be used to compute the expected number of molecules in each state, denoted \( E[Y_{S_i}] \). These expectations reflect how signal intensity propagates through the cascade and allow one to compute the corresponding Fisher Information and ultimately the system’s information capacity.

\noindent For each state \(S_i\) (\(i = 1, 2, \ldots, m\)), the rate of change of the expected number of molecules \(E[Y^i]\) is zero, therefore at steady state

 \[ 
 E[Y_{S_1}] = \frac{\beta_0+\kappa_0^{\max} \cdot h(x)}{\gamma + \kappa}, 
 \]
 
 \[
 E[Y_{S_2}] = \frac{k E[Y_{S_1}]}{\gamma + \kappa} = \frac{(\beta_0+\kappa_0^{\max} \cdot h(x)) \cdot \kappa}{(\gamma + \kappa)(\gamma + \kappa)},
 \]
 for  $i = 1, 2, \ldots, m-1$
 \[
 E[Y_{S_i}] = \frac{ (\beta_0+\kappa_0^{\max} \cdot h(x) )\cdot \kappa^{i-1}}{(\gamma + \kappa)^i},  \]
 and for the last $m$-th state
 \[
 E[Y_{S_m}] = \frac{ (\beta_0+\kappa_0^{\max} \cdot h(x) )\cdot \kappa^{m-1}}{\gamma(\gamma + \kappa)^{m-1}}. 
 \]

\noindent
We have therefore
\[
Y_{S_m} \sim \text{Poisson}\left( \frac{\beta_0 + \kappa_0^{\max} \cdot h(x)}{\gamma} \cdot \alpha^{m-1} \right),
\]
where
\[
\alpha = \left( \frac{\kappa}{\kappa + \gamma} \right),
\]
and, using Eq.~\ref{eq:V},
\[
V = \int_{x_{\min}}^{x_{\max}} \frac{1}{\sqrt{ \frac{1}{\gamma} (\beta_0 + k_0^{\max} \cdot h(x)) \cdot \alpha^{m-1} }} 
\cdot \left| \frac{\mathrm{d}}{\mathrm{d}x} \left( \frac{1}{\gamma} k_0^{\max} \cdot \alpha^{m-1} \cdot h(x) \right) \right| \, \mathrm{d}x
= \int_{\lambda_{\min}}^{\lambda_{\max}} \frac{1}{\sqrt{\lambda}}\,\mathrm{d}\lambda,
\]
\[
= 2 \sqrt{ \frac{\alpha^{m-1}}{\gamma} (\beta_0 + k_0^{\max} \cdot h(x_{\max})) }
- 2 \sqrt{ \frac{\alpha^{m-1}}{\gamma} (\beta_0 + k_0^{\max} \cdot h(x_{\min})) }.
\]

\noindent
Assuming \( h(x_{\min}) = h(0) = 0 \) and \( h(x_{\max}) = h(\infty) = 1 \), we obtain
\[
V = 2 \alpha^{(m-1)/2} \left( \sqrt{ \frac{\beta_0 + \kappa_0^{\max}}{\gamma} } - \sqrt{ \frac{\beta_0}{\gamma} } \right),
\]
and therefore the asymptotic information capacity is
\[
C^*_{A} = \log_2 \left( \frac{2 \alpha^{(m-1)/2}}{(2\pi e)^{1/2}} 
\left( \sqrt{ \frac{\beta_0 + \kappa_0^{\max}}{\gamma} } - \sqrt{ \frac{\beta_0}{\gamma} } \right) \right).
\]

\noindent
\noindent
To make explicit how information capacity depends on cascade length and kinetic parameters, it is helpful to rewrite the expression as
\begin{equation}
C^*_{A} = \log_2\left( \frac{2}{(2\pi e)^{1/2}} \left( \sqrt{ \frac{\beta_0 + \kappa_0^{\max}}{\gamma} } - \sqrt{ \frac{\beta_0}{\gamma} } \right) \right) 
+ \frac{1}{2}(m - 1) \log_2\left( \frac{\kappa}{\kappa + \gamma} \right).
\end{equation}

\noindent 
The resulting expression provides a quantitative account of an inherent limitation in multistep signaling cascades: as a signal is transmitted through successive biochemical states, information is inevitably lost. 
This decline is not merely plausible, it is mandated by the data processing inequality, a foundational result in information theory \cite{cover2012elements}, which guarantees that post-processing steps cannot increase the information about the input. Here, that principle is made explicit: the derived formula directly links molecular parameters, conversion rate \( \kappa \), degradation rate \( \gamma \), and cascade length \( m \), to an upper bound on information transfer, exposing the fundamental constraints imposed by network topology and kinetics. The  capacity decreases linearly with the number of steps \( m \), modulated by the ratio \( \frac{\kappa}{\kappa + \gamma} < 1 \), which encodes the trade-off between forward progression and degradation. The formula reflects the cumulative effect of biochemical noise introduced at each stage. Figure~\ref{fig:cascade}B illustrates this behavior: longer cascades or increased degradation result in greater information loss, progressively limiting the system’s ability to resolve input values. While this model deliberately abstracts away from biological complexities such as temporal averaging, enzyme saturation, spatial effects, or feedback regulation, it serves as a transparent and analytically tractable benchmark. It demonstrates how the Fisher Information formalism can reveal structural trade-offs in signaling design and identify limits on information fidelity imposed purely by network architecture and stochastic kinetics.

\clearpage

\subsection{Gaussian distribution}
\noindent
The Gaussian (normal) distribution is widely used to model biological signaling systems with 
continuous outputs, especially when the system integrates many  stochastic molecular events 
(e.g., transcriptional bursts, receptor activation, or diffusion) that, in aggregate, lead 
to approximately normal response distributions. Examples include protein or mRNA concentrations 
in single cells, intracellular calcium levels, or nuclear localization of signaling factors, 
where the mean response and its variability are continuous-valued and often experimentally accessible. 

\noindent
Here, we apply the Fisher Information framework to systems with Gaussian-distributed outputs and consider two biologically motivated cases: (1) a constant variance model, and (2) a signal-dependent variance model. The first reflects situations where variability remains steady across input conditions, e.g., when noise is dominated by baseline  or other signal independent fluctuations. The second captures the widespread biological observation that noise often scales with the mean, as seen in gene expression and cytokine signaling \cite{paulsson2004summing, raser2005noise}. These cases allow us to evaluate how different noise models impact the information capacity of signaling systems and how this capacity changes with signal  range, signal fold-change and cell population size.

\noindent
The FI for the Gaussian distribution, $Y \sim \mathcal{N}(\mu, \sigma^2)$ with the mean $\mu$ and variance $\sigma^2$
and the probability density function 
\[
f(x; \mu, \sigma^2) = \frac{1}{\sqrt{2 \pi \sigma^2}} \exp\left(-\frac{(x - \mu)^2}{2\sigma^2}\right).
\]
is given as 
\[
FI(\mu, \sigma^2) = 
\begin{bmatrix}
\frac{1}{\sigma^2} & 0 \\
0 & \frac{1}{2\sigma^4}
\end{bmatrix}.
\]

\subsubsection*{Special Case: fixed \( \sigma \)}
\noindent  The first case is where the mean $\mu$ varing from $\mu_{\min}$ to $\mu_{\max}$ is the input signal, while the variance \( \sigma^2 \) remains constant. Using transformation rule given by Eq. \ref{eq:FiTrans}, computation of FI with respect to $\mu$ is straightforward as $\frac{\partial \sigma^2}{\partial \mu}=0$. We have

$$FI(\mu)=\frac{1}{\sigma^2}.$$

\noindent
Therefore, as defined in Eq. \ref{eq:V}, we have  
\[
V = \frac{\mu_{\max} -\mu_{\min}}{\sigma},
\]
which here takes the form of the signal-to-noise-ratio (SNR) commonly used in engineering. The asymptotic information capacity of the system is then

\[
C^*_{A} = \log_2\left(\frac{\mu_{\max} - \mu_{\min}}{(2\pi e)^{\frac{1}{2}} \sigma } \right).
\]
\noindent
This example formalizes the classical concept of SNR within 
the information-theoretic framework of information capacity. In the case of constant variance, the asymptotic information capacity depends logarithmically on the dynamic range of the mean response divided by the noise level providing a quantitative link between intuitive engineering heuristics and a formally derived capacity.

\subsubsection*{Special Case: \( \sigma = \lambda \mu \)}
\noindent Now consider a scenario where the variability of the response increases proportionally to the mean. This corresponds to biological systems where the strength of noise grows linearly with the signal, which is often exactly or approximately observed \cite{taniguchi2010quantifying, newman2006single, paulsson2004summing, topolewski2022phenotypic}. We model this by assuming \( \sigma^2 = (\lambda \mu)^2 \), where \( \lambda \) is a constant proportionality factor. In this case, we have that $\frac{\partial \sigma^2}{ \partial \mu }=2\lambda^2\mu$, therefore the Fisher Information for the mean \( \mu \) is obtained using a transformation Jacobian, Eq. \ref{eq:FiTrans}, $J=[1,2\lambda^2\mu]^T$,

\[
FI(\mu) = \frac{1}{\mu^2}\left( \frac{1}{\lambda^2} + 2 \right),
\]

\noindent leading to $V$ defined in Eq. \ref{eq:V} given as

\[
V = \sqrt{\frac{1}{\lambda^2} + 2} \left( \ln\left(\frac{\mu_{\max}}{\mu_{\min}}\right)\right).
\]

\noindent The corresponding asymptotic capacity is

\begin{equation}\label{eq:Gaussian:CA}
C^*_{A} = \log_2\left( \frac{1}{(2\pi e)^{\frac{1}{2}}} \sqrt{\frac{1}{\lambda^2} + 2} \left(\ln\left(\frac{\mu_{\max}}{\mu_{\min}}\right)\right) \right).
\end{equation}

\noindent This case demonstrates how scaling of the variability with signal level, a feature often observed in biological systems such as cytokine signaling or gene expression leads to information capacity being predominantly dependent on fold-change of the mean response ${\mu_{\max}}/{\mu_{\min}}$, as opposed to additive increase in the constant variance scenario. 

\subsubsection*{Example: Single Cell Signaling}
\noindent Now consider a concrete biological example that helps to interpret the formulas above. 
Let \( x \) denote a ligand concentration (such as a cytokine), and let \( Y^{(i)} \) be the response 
of an individual cell \( i \), e.g., the nuclear concentration of a transcription factor. 
We model this response as normally distributed with mean \( \mu \) and standard deviation \( \sigma \) 
that is proportional to the mean, so that \( \sigma^2 = (\lambda \mu)^2 \). Assuming that each cell has an identical mean response function \( h(x) \), the asymptotic information capacity can be computed by substituting $\mu_{max}=h(x_{\max})$ and $\mu_{min}=h(x_{\min})$ into Eq. \ref{eq:Gaussian:CA},
which gives the information capacity regarding ligand concentration stored jointly in \( N \) cells
\begin{equation}
C^*_{N} = \log_2\left( \frac{1}{(2\pi e)^{\frac{1}{2}}} \sqrt{\frac{1}{\lambda^2} + 2} \left( \ln\left( \frac{h(x_{\max})}{h(x_{\min})} \right) \right) \right)
+ \frac{1}{2} \log_2(N).
\end{equation}
\noindent 
This result highlights two  insights. First, when noise scales with the signal, the system’s capacity 
to resolve input levels is governed primarily by the fold-change in the output response, not the 
absolute range, supporting the idea that many signaling systems operate in a fold-change detection mode 
\cite{cohen2011dynamic, goentoro2009evidence, nienaltowski2021fractional}. Second, pooling responses 
from multiple cells increases the capacity additively in \( \frac{1}{2}\log_2(N) \), reflecting classic results 
from statistical estimation theory. 

 \subsection{Gamma Distribution}
 
 \noindent The Gamma distribution is another relevant model for biological systems, especially when the data is skewed or limited to positive values. The probability density function of a Gamma-distributed random variable \(Y\) with shape parameter \(\alpha\) and rate parameter \(\beta\) is given by
\[
f(y; \alpha, \beta) = \frac{\beta^\alpha y^{\alpha-1} e^{-\beta y}}{\Gamma(\alpha)}, \quad y > 0,
\]
where \(\alpha > 0\) and \(\beta > 0\). The mean \(\mu\) and variance \(\sigma^2\) are related to \(\alpha\) and \(\beta\) as
\[
\mu = \frac{\alpha}{\beta}, \quad \sigma^2 = \frac{\alpha}{\beta^2}.
\]
Thus, the shape and rate parameters can be reparametrized in terms of \(\mu\) and \(\sigma^2\) as
\[
\alpha = \frac{\mu^2}{\sigma^2}, \quad \beta = \frac{\mu}{\sigma^2}.
\]
The FI in terms of \((\alpha, \beta)\) is
\[
FI(\alpha, \beta) = 
\begin{bmatrix}
\psi'(\alpha) & -\frac{1}{\beta} \\
-\frac{1}{\beta} & \frac{\alpha}{\beta^2}
\end{bmatrix},
\]
where \(\psi'(\alpha)\) is the trigamma function, the derivative of the digamma function \(\psi(\alpha)\).
Using the variable transformation, Eq. \ref{eq:FiTrans}, from \((\alpha, \beta)\) to \((\mu, \sigma^2)\),  we have the Jacobian matrix of the transformation
\[
J = 
\begin{bmatrix}
\frac{\partial \alpha}{\partial \mu} & \frac{\partial \alpha}{\partial \sigma^2} \\
\frac{\partial \beta}{\partial \mu} & \frac{\partial \beta}{\partial \sigma^2}
\end{bmatrix} = 
\begin{bmatrix}
\frac{2\mu}{\sigma^2} & -\frac{\mu^2}{(\sigma^2)^2} \\
\frac{1}{\sigma^2} & -\frac{\mu}{(\sigma^2)^2}
\end{bmatrix}
\]
resulting in FI with respect \(\mu\) and \(\sigma^2\) 
\[
FI(\mu, \sigma^2) =
\begin{bmatrix}
\frac{\sigma^2}{\mu^2} \psi' \left( \frac{\mu^2}{\sigma^2} \right) & -\frac{\sigma^4}{\mu^3} \psi' \left( \frac{\mu^2}{\sigma^2} \right) \\
-\frac{\sigma^4}{\mu^3} \psi' \left( \frac{\mu^2}{\sigma^2} \right) & \frac{\sigma^6}{\mu^4} \psi' \left( \frac{\mu^2}{\sigma^2} \right) + \frac{2\sigma^4}{\mu^2}
\end{bmatrix}.
\]
Although the above formula is useful for numerical studies its manual manipulation is difficult. We focus, therefore, similarly to the gaussian scenario, on a special case. Here, however, only on the case where $\sigma=\lambda \mu$ as the case of constant scenario does not seem to have a simple form.
\subsubsection*{Special case: $\sigma=\lambda \mu$}
\noindent In the considered special case, we have that
\[
\begin{aligned}
\alpha &= \frac{1}{\lambda^2} \quad \beta = \frac{1}{\lambda^2 \mu},
\end{aligned}
\]
therefore, 
\[
\begin{aligned}
\frac{\partial \alpha}{\partial \mu}&=0 \quad \frac{\partial \beta}{\partial \mu}= -\frac{1}{\lambda^2 \mu^2}
\end{aligned}
\]
and $J = [0, -\frac{1}{\lambda^2 \mu^2}]^T.$
After multiplication $J^T FI(\alpha, \beta) J$, we get Fisher Information with respect to $\mu$
\[FI(\mu)=\frac{1}{\lambda^2 \mu^2}\]
and
\[ V=\left( \frac{1}{\lambda} \right) \left(\ln \left(\frac{ \mu_{max}}{ \mu_{min} }\right)\right). \] The capacity is then given as
\begin{equation}
C^*_{A}= \log_2\left(\frac{1}{(2\pi e)^{\frac{1}{2}}} \left( \frac{1}{\lambda} \right) \left(\ln \left(\frac{ \mu_{max}}{ \mu_{min} }\right)\right) \right).
\end{equation}
Similarly as in the Gaussian scenario, the assumption $\sigma=\lambda \mu$ implies dependance of the asymptotic capacity on the fold-change of the mean. 
\subsection*{Example: Single cell signaling}
\noindent 
The formula can be used in the same way as in the Gaussian example to study signaling capacity of $N$ cells. Considering the same example, for outputs distributed as Gamma we have the following capacity
\begin{equation}
C^*_{N}= \log_2\left(\frac{1}{(2\pi e)^{\frac{1}{2}}} \left( \frac{1}{\lambda} \right) \left(\ln \left(\frac{ h(x_{max})}{ h(x_{min}) }\right)\right) \right)
 + \frac{1}{2}\log_2(N).
\end{equation}
\pagebreak

\subsection{Information Capacity in Heterogeneous Cell Populations}

\noindent
Having derived closed-form expressions for information capacity in canonical settings, we now apply them to a more complex, biologically relevant scenario: a heterogeneous cell population 
\cite{wada2020single, topolewski2022phenotypic}. We assume the population comprises \( m \) distinct phenotypes, denoted as \( Z \in \{1,\dots, j,  \dots, k\} \). Each cell independently adopts phenotype \( j \) with probability \( \omega_j \), such that \( \sum_{j=1}^k \omega_j = 1 \). The conditional probability of observing response \( Y \) from a cell of phenotype \( j \), given input \( X = x \), is denoted as \( P_j(Y \mid X = x) \). The joint distribution of \( (Y, Z) \) given \( x \) is

\[
P(Y, Z = j \mid X = x) = \omega_j P_j(Y \mid X = x).
\]

\noindent
The Fisher Information with respect to \( x \) for this mixture model is given by

\[
FI(x) = \mathbb{E} \left[ \left( \frac{\partial}{\partial x} \log P(Y, Z \mid X = x) \right)^2 \right].
\]

\noindent
Expanding the expectation over phenotype assignments and conditioning on \( Z = j \), we obtain

\[
FI(x) = \sum_{j=1}^k \omega_j \, \mathbb{E}_{Y \sim P_j} \left[ \left( \frac{\partial}{\partial x} \log P_j(Y \mid X = x) \right)^2 \right] = \sum_{j=1}^k \omega_j \, FI_j(x),
\]

\noindent
where \( FI_j(x) \) is the Fisher Information for phenotype \( j \). The integrated square root of the total Fisher Information becomes

\[
V = \int_{x_{\min}}^{x_{\max}} \sqrt{ \sum_{j=1}^k \omega_j FI_j(x) } \, dx.
\]

\noindent
In the special case, where the Fisher Information is localized, that is, each \( x \) activates only one phenotype significantly (i.e., \( FI_j(x) \gg 0 \) for a single \( j \)), the integral simplifies to

\[
V = \sum_{j=1}^k \sqrt{ \omega_j } V_j, \quad \text{where} \quad V_j = \int_{x_{\min}}^{x_{\max}} \sqrt{ FI_j(x) } \, dx.
\]

\noindent
These expressions give opportunity us to reuse the analytical results derived earlier, as each \( V_j \) corresponds to the integrated Fisher Information of a specific distribution. Assuming all phenotypes share the same integrated Fisher Information (i.e., \( V_j = V_0 \) for all \( j \)), and are equiprobable (i.e., \( \omega_j = \frac{1}{k} \)), we obtain

\begin{equation}
V = \sqrt{k} \cdot V_0
\end{equation}

and
\begin{equation}
C^*_N = \log_2 \left( \frac{1}{(2\pi e)^{1/2}} V_0 \right) + \frac{1}{2} \log_2(kN).
\end{equation}

\noindent This final expression reveals that in the limit of non-overlapping phenotypic responses, the sensing capacity of a heterogeneous cell population scales as $\frac{1}{2}\log_2(kN)$, where $N$ is the number of cells and 
$m$ the number of distinct response phenotypes. This scaling implies that phenotypic diversity can enhance population-level information transfer, essentially by $\frac{1}{2}\log_2(k)$ when phenotypes are well-separated. Crucially, the gain is independent of the specific form of the response distribution and arises without increasing single-cell precision, just by partitioning the input range across cell types.

\noindent
To use concrete distributional forms, suppose the response of phenotype \( j \) has a mean \( \mu_j(x) = h_j(x) \), and standard deviation \( \sigma_j(x) = \lambda \mu_j(x) \), where \( \lambda \) is a constant.  Assume each response function spans the range of output values

\[
h_j(x_{\min}) = \mu_{\min}, \qquad h_j(x_{\max}) = \mu_{\max}, \quad \text{for all } j,
\]
however for different ranges of input values, ensuring the assumption of localized Fisher information is satisfied.
\noindent
Then, using our earlier derivations, the information capacity for \( N \) cells and \( m \) phenotypes
for Gaussian-distributed responses is given by

\begin{equation} \label{eq:C_N:k_phenotypes_Gaussian_cells}
C^*_N = \log_2 \left( \frac{1}{(2\pi e)^{1/2}} \sqrt{ \frac{1}{\lambda^2} + 2 } \cdot \ln \left( \frac{\mu_{\max}}{\mu_{\min}} \right) \right) + \frac{1}{2} \log_2(kN),
\end{equation}
and for Gamma-distributed responses by
\begin{equation} \label{eq:C_N:k_phenotypes_Gamma_cells}
C^*_N = \log_2 \left( \frac{1}{(2\pi e)^{1/2}} \cdot \frac{1}{\lambda} \cdot \ln \left( \frac{\mu_{\max}}{\mu_{\min}} \right) \right) + \frac{1}{2} \log_2(kN).
\end{equation}

\noindent
The above formulas quantify the sensing capacity of a population of \( N \) cells.
with \( m \) distinct cellular phenotypes in the limit of non-overlapping responses distributed according to Gaussian and Gamma distribution. 

\begin{figure}[!h]
\centering
\begin{minipage}{\linewidth}
 \flushleft
% {\bf (A)}\\ % Label for the first figure
 \includegraphics[width=\linewidth]{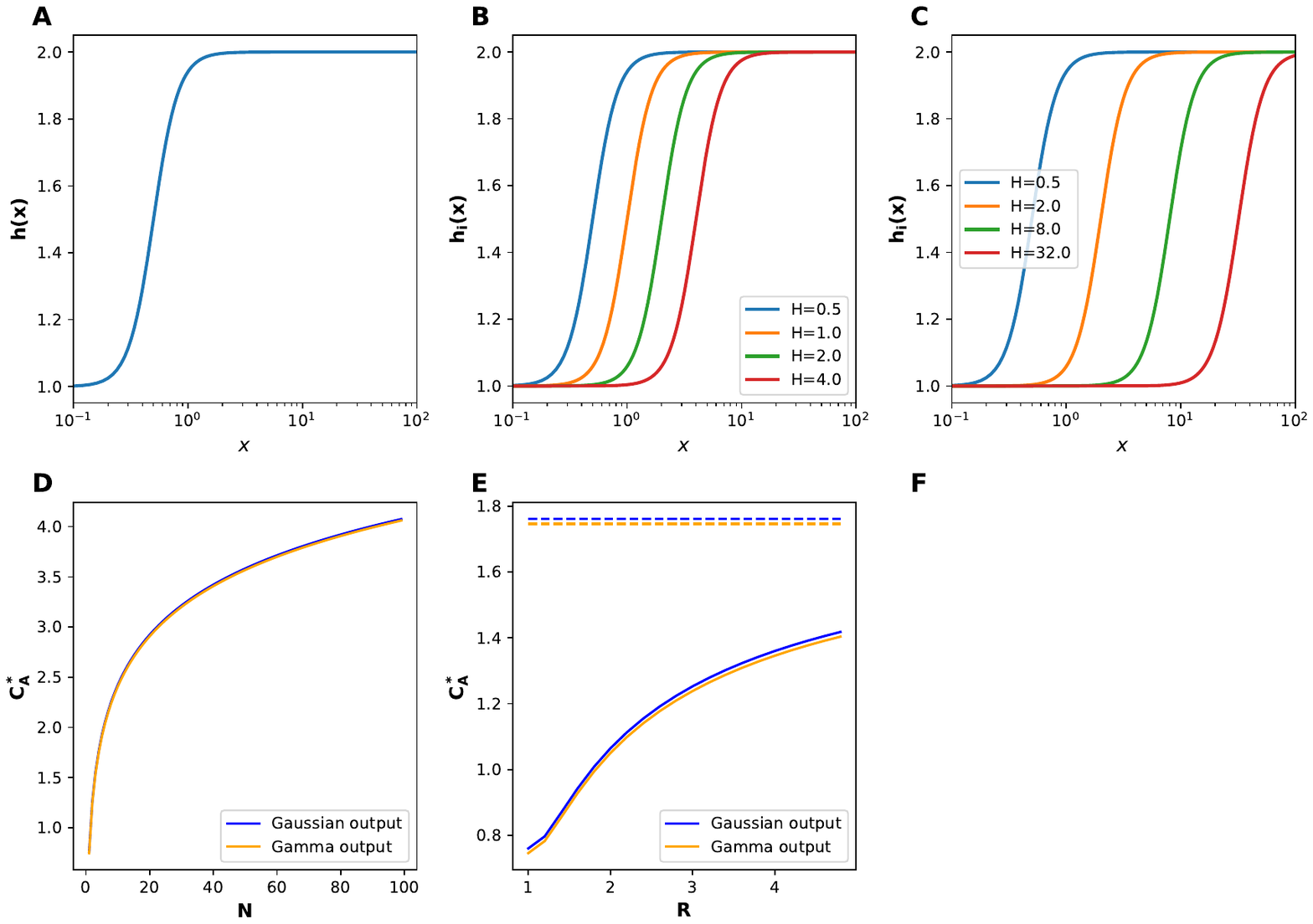}
\end{minipage}
\caption{\ \\
{\bf (A)} Mean response of a cell in a scenario where all cells are governed by the same response parameters, and mean response is described by $h(x)=\mu_{min}+ (\mu_{max}-\mu_{min})\frac{(x/H)^n}{1+(x/H)^n}$ with $\mu_{min}=1$, $\mu_{max}=2$, $H=1$, $n=2$.\\
{\bf (B)} Mean responses of four, $k=4$, distinct phenotypes with strongly overlapping sensitivity ranges. Mean response is modeled as in scenario (A) with phenotypes having different values of $H$: $H_1=0.5$, $H_2=1$, $H_3=2$, $H_4=4$, and hence $R=2$.\\
{\bf (C)} As in (B) but for phenotypes with small overlapping sensitivity ranges: $H_1=0.5$, $H_2=2$, $H_3=8$, $H_4=32$, and hence $R=4$.\\
{\bf (D)} Capacity $C_N$ as a function of $N$ in a scenario with homogenous mean response, as in (A) with Gaussian and Gamma output, for $\sigma=\lambda \mu$ and $\lambda=0.1$.\\
{\bf (E)} Capacity $C^*_A$ in a scenario with four response phenotypes $(m=4)$ as a function of the ratio $R=H_{i+1} / H_{i}$, for Gaussian and Gamma output, $\sigma=\lambda \mu$ and $\lambda=0.1$, together with the limit of fully distinct responses established by Eq. \ref{eq:C_N:k_phenotypes_Gamma_cells} and Eq. \ref{eq:C_N:k_phenotypes_Gamma_cells}, plotted as dashed lines.
}
\label{fig:cells}
\end{figure}

\noindent
We further examine how it depends on overlaps between response sensitivities of different phenotypes by considering a numerical example in which each phenotype \( j \) exhibits a distinct mean response function defined as
\[
h_j(x) = \mu_{\min} + (\mu_{\max} - \mu_{\min}) \cdot \frac{(x/H_j)^n}{1 + (x/H_j)^n},
\]
where \( \mu_{\min} \) and \( \mu_{\max} \) are the minimum and maximum possible response levels, respectively, Fig \ref{fig:cells}A. The Hill function parameter \( H_j \) controls the input value at which phenotype \( j \) reaches half its maximum activation, and \( n \) determines the steepness of the response. Further, similarly as in the allosteric sensor example, we define the ratio \( R = H_{j+1}/H_j \), which quantifies the spacing between the sensitivity ranges of adjacent phenotypes. As \( R \) increases, the activation curves of different phenotypes become more distinct and less overlapping, approaching the idealized limit in which each phenotype operates over a localized region of input space, Fig \ref{fig:cells}B-C. In this limit, the system approaches the maximum capacity predicted by Eq.~\ref{eq:C_N:k_phenotypes_Gaussian_cells} for the Gaussian case and Eq.~\ref{eq:C_N:k_phenotypes_Gamma_cells} for the Gamma case, Fig \ref{fig:cells}D-E.

\noindent
This example also underscores the practical value of the Fisher-based formalism. The explicit capacity formulas enable a decomposition of how population size, phenotypic heterogeneity, noise scaling, and dynamic range each contribute to information transfer. This analytic clarity facilitates quantitative assessment of trade-offs in sensing system design. Whether the goal is to understand adaptive advantages of diversity in natural populations or to optimize synthetic biosensors, the framework here identifies when heterogeneity enhances performance, and when it does not.

\clearpage
\section{Discussion}

\noindent
The analytical expressions we derive for the asymptotic information capacity \( C^*_A \), summarized in Table~\ref{tab:asymptotic_capacities}, offer a practical toolset for evaluating the limits of signal resolution in biological systems. By expressing capacity in terms of Fisher Information, our framework reveals how type of output distribution, kinetic rates, input values range, input-dependent noise, and population heterogeneity, quantitatively shapes the ability of a system to encode and transmit information. Unlike conventional approaches based on numerical mutual information estimation (e.g., via Blahut–Arimoto or Monte Carlo sampling), our method yields closed-form results or efficiently computable integrals that directly expose how parameters control capacity. In particular, we show that for a wide class of biologically relevant distributions, Binomial, Multinomial, Poisson, Gaussian, and Gamma, the capacity can be expressed as a scaled log of a resolved input volume \( V \), integrating the local Fisher Information over input space. This links information transmission directly to geometric properties of the statistical model, and establishes a principled, computation-free approximation of capacity in large-$N$ systems.

\medskip
\noindent
One interesting insight is the quantification of how structural features such as noise scaling alter the dependence of capacity on signal range. When noise scales with the mean (as in Gamma or scaled-Gaussian models), capacity depends on the \textit{fold-change} in signal levels, a logarithmic dependence consistent with fold-change detection observed across diverse signaling systems. Conversely, when noise is additive and fixed (constant-variance Gaussian), capacity scales with the absolute signal range, i.e., signal-to-noise ratio. This provides formula based  explanation of why biological sensors often encode relative changes, rather than absolute concentrations.

\medskip
\noindent
We also show that phenotypic heterogeneity, when organized into non-overlapping sensitivity regimes, contributes additively to capacity. In such cases, the capacity scales as \( \frac{1}{2} \log_2(mN) \), where \( m \) is the number of distinct phenotypes and \( N \) the number of cells. This allows direct quantification of the benefit of diversity, offering a design principle for synthetic circuits or a testable hypothesis for natural systems: if subpopulations are tuned to distinct input regimes, the net sensing capacity increases without altering the precision of individual cells.

\medskip
\noindent
Finally, our analysis of signal degradation in multi-step cascades reveals a fundamental constraint: information decays linearly with pathway length. The derived expressions isolate how degradation rate and conversion efficiency impose strict upper bounds on usable capacity. While based on idealized assumptions, these formulas provide analytic limits that any realistic system must obey, offering useful benchmarks for both biological modeling and synthetic design.

\medskip
\noindent
In summary, the value of this framework is threefold: (i) it replaces costly optimization with tractable integrals; (ii) it exposes how distributional assumptions and parameters impact resolvability; and (iii) it enables decomposition of sensing performance into interpretable components, range, noise, population, and dynamics. Beyond biology, the same principles apply to any noisy measurement system, sensor networks, data compression, or analog-to-digital conversion, where understanding the structure–capacity relationship is essential.

\begin{table}[!h]
\centering
\caption{Summary of Derived Formulas for Asymptotic Information Capacity \( C^*_A \)}
\resizebox{\textwidth}{!}{
\begin{tabular}{|l|l|p{4cm}|l|}
\hline
\textbf{Distribution} & \textbf{Parameters}    & \textbf{Additional Assumptions}        & \textbf{Asymptotic Capacity \( C^*_A \)}         \\ \hline
Binomial    & $p_{\text{min}}, p_{\text{max}}$ & -           & $C^*_A = \log_2 \left( (2\pi e)^{-1/2} (\arcsin(2p_{\text{max}}-1) - \arcsin(2p_{\text{min}}-1)) \right)$ \\ \cline{2-4}
      & $p_{\text{min}}, p_{\text{max}}$ & $p_{\text{min}} = 0, p_{\text{max}} = 1$      & $C^*_A = \log_2 \left( (2\pi e)^{-1/2} \pi \right)$ \\ \hline
Multinomial   & $k, \omega_i, h_i(x)$    & Non-overlapping sensitivity ranges    & $C^*_A = \frac{1}{2}\log_2 \left( \frac{k \cdot \pi}{2e/\pi} \right)$          \\ \hline
Poisson    & $\lambda_{\text{min}}, \lambda_{\text{max}}$ & large    $\lambda_{\text{min}}$    & $C^*_A = \log_2 \left( (2\pi e)^{-1/2} (2\sqrt{\lambda_{\text{max}}} - 2\sqrt{\lambda_{\text{min}}}) \right)$ \\ \hline
Gaussian    & $\mu_{\text{min}}, \mu_{\text{max}}, \sigma$ & Fixed $\sigma$          & $C^*_A = \log_2 \left( \frac{\mu_{\text{max}} - \mu_{\text{min}}}{(2\pi e)^{1/2} \sigma} \right)$ \\ \cline{2-4}
      & $\mu_{\text{min}}, \mu_{\text{max}}, \lambda$ & $\sigma = \lambda \mu$         & $C^*_{A} = \log_2\left( \frac{1}{(2\pi e)^{\frac{1}{2}}} \sqrt{\frac{1}{\lambda^2} + 2} \left(\ln\left(\frac{\mu_{\max}}{\mu_{\min}}\right)\right) \right)$
 \\ \hline
Gamma     & $\mu_{\text{min}}, \mu_{\text{max}}, \lambda$ & $\sigma = \lambda \mu$          & $C^*_{A}= \log_2\left(\frac{1}{(2\pi e)^{\frac{1}{2}}} \left( \frac{1}{\lambda} \right) \left(\ln \left(\frac{ \mu_{max}}{ \mu_{min} }\right)\right) \right)$
 \\ \hline
\end{tabular}
}
\label{tab:asymptotic_capacities}
\end{table}
\clearpage

\section{Appendix}
This appendix provides additional detail on the implementation of our framework. This included a step-by-step algorithmic outline and its connection to existing  small-noise approximation described in \cite{tkavcik2008information}. We also provde a link to a python code to reproduce computations for Figures presented in the maniscript.

\subsection{Step-by-step Algorithmic Outline}
In order to evaluate capacity of a given system the following steps should be followed.
\begin{enumerate}
    \item \textbf{Define the Input-Output Model.} Specify $P(Y\mid X=x)$, which captures
           biochemical network (e.g., receptor binding, enzymatic reactions, transcription
          factor dynamics). For multi-input, multi-output or time-resolved scenarios, $x \in \mathbb{R}^l$
          and $y \in \mathbb{R}^m$ may be multi-dimensional.
\item \textbf{Derive/Compute $\mathrm{FI}(x)$.} 
 If $P(Y \mid X = x)$ belongs to a known family (e.g., binomial, Poisson, or Gamma), the Fisher Information 
   often has closed-form expressions, of which most common and presumably useful distributions are present in Table \ref{tab:asymptotic_capacities}.  Otherwise, one typically needs to take partial derivatives 
   of $\log P(Y\mid X=x)$, which can become challenging. In such cases, numerical approaches are advisable.  Several software packages or computational approaches can facilitate derivations and integrals \cite{sheppard2013spsens,komorowski2012stochsens,hu2018isap, lakatos2015multivariate}.  Other more general techniques also exist for empirical or nonparametric estimation of Fisher Information \cite{berisha2015empirical,har2016nonparametric}.

    \item \textbf{Integrate over $x$.} Evaluate $V = \int_{\mathcal{X}} \sqrt{\vert \mathrm{FI}(x)\vert}\,\mathrm{d}x$ 
          numerically. If a model utilizes the log-scale parameterization $\mathrm{FI}(x)$ is multiplied by 
          $(x\,\ln(10))^2$.  
    \item \textbf{Compute $C_{\mathrm{A}}^*$.} Substitute $V$ into Eq.~\ref{eq:CapacityApp} to
          find the asymptotic capacity. For $N>1$, use Eq.~\ref{eq:CapacityN} to estimate
          $C_N^*$.

    \item \textbf{Interpret Negative $C_{\mathrm{A}}^*$ if it arises.}
          This can occur if $\vert \mathrm{FI}(x)\vert$ is small over $\mathcal{X}$, meaning
          the system’s outputs barely distinguish small perturbations in $x$. While $C_{\mathrm{A}}^*$ 
          can be negative, the joint capacity $C_N^*$ still grows to reach positive values for sufficiently large  $N$. If the goal is to get accurate very estimates of the single-receiver capacity $C_1^*$ and $C^*_A$ turns out to be negative a direct approach, such as numerical integration or Blahut--Arimoto is required.\end{enumerate}

\subsection{Comparison to Small-Noise Approximations (SNA)}
\noindent 
Many theoretical analyses of transcriptional regulation leverage a small-noise approximation (SNA),
particularly in models where the system’s output $Y$ at each input $x$ is well-described by a
Gaussian distribution of relatively small variance $\sigma^2(x)$. We briefly summarize a derivation of this approach, detailed by Tka\v{c}ik \emph{et al.} \cite{tkavcik2008information}.

\noindent 
If $P(Y \mid x)$ is narrowly peaked around its mean $\mu(x)$,
then we can write
\begin{equation}\label{eq:normal_tkacik}
  P(y \mid x) \;\approx\;
    \frac{1}{\sqrt{2\pi\,\sigma^2(x)}}\,
    \exp\!\Bigl[
      -\,\frac{\bigl(y - \mu(x)\bigr)^2}
            {2\,\sigma^2(x)}
         \Bigr].
\end{equation}
For each input $x$, if the noise level $\sigma(x)$ is small relative to the mean $\mu(x)$ than
 the mutual information $\mathrm{MI}(X,Y)$ can be expanded in powers of the noise amplitude
and the information capacity formula can be derived \cite{tkavcik2008information},
\begin{equation}
\label{eq:TkaEq8-main}
  C^*(X,Y)\approx \log_2\left( \frac{ Z}{\sqrt{ 2\pi e}}\right),
 \end{equation}
where 
\begin{equation} 
Z=\int_{\mathcal X}\frac{\partial \mu/ \partial x }{ \sigma(x)}dx.
 \end{equation}
\noindent 
For comparison, consider the Fisher information of the above model. The Fisher Information for the Gaussian distribution, Eq. \ref{eq:normal_tkacik}, is given as
\[
FI(x)=\left( \frac{\partial \mu/ \partial x }{ \sigma(x)}\right)^2+\frac{1}{2} \frac{ (\partial \sigma^2(x)/ \partial x)^2 }{\sigma^4(x)}
\]
By retaining only the first-order approximation and beglecting the contribution of variance dependent on the signal, x, we have 
\[
\left(
\tfrac{\partial\mu/\partial x}{\sigma(x)}
\right)^2.\]

Substituting the above into Eqs.~\ref{eq:V} and 
\ref{eq:CapacityApp} reproduces Eq. \ref{eq:TkaEq8-main}, thus clarifying that SNA 
is effectively a special case within FI-based framework.

\subsection{Code repository}
\noindent
Figures presented in the manuscript have been generated using Python scrips deposited under the following link:\\
\url{https://github.com/sysbiosig/Capacity-of-Canonical-Models }

\section*{Funding}
This research was supported by the National Science Center, Poland, under grant number 2020/39/B/NZ2/03259.

\bibliography{sample}

\end{document}